\documentclass[reprint,amsmath,amssymb,aps,prl,superscriptaddress,preprintnumbers]{revtex4-1}

\usepackage{graphicx}
\usepackage{dcolumn}
\usepackage{bm}
\usepackage{hyperref}
\usepackage{lineno}

\usepackage{multirow}
\usepackage{enumitem}
\usepackage[english]{babel}
\usepackage{siunitx}
\usepackage{times}
\usepackage{verbatim}
\usepackage{amssymb}

\usepackage[normalem]{ulem}
\usepackage{accents}
\usepackage[tight]{units} 

\newcommand*{\pbar}[1]{\accentset{(-)}{#1}}

\newcommand {\cls}{\ensuremath{\rm{CL_{\it s}}}}

\usepackage{color}
\usepackage{xspace}

\begin{document}

\preprint{FERMILAB-PUB-20-054-ND}

\title{Improved Constraints on Sterile Neutrino Mixing from Disappearance Searches in the
MINOS, MINOS+, Daya Bay, and Bugey-3 Experiments}

%
\newcommand{\NUU}{{National~United~University, Miao-Li}}
\newcommand{\Minnesota}{University of Minnesota, Minneapolis, Minnesota 55455, USA}
\newcommand{\ECUST}{{Institute of Modern Physics, East China University of Science and Technology, Shanghai}}
\newcommand{\SJTU}{{Department of Physics and Astronomy, Shanghai Jiao Tong University, Shanghai Laboratory for Particle Physics and Cosmology, Shanghai}}
\newcommand{\SLAC}{Stanford Linear Accelerator Center, Stanford, California 94309, USA}
\newcommand{\JMU}{Physics Department, James Madison University, Harrisonburg, Virginia 22807, USA}
\newcommand{\Delhi}{Department of Physics \& Astrophysics, University of Delhi, Delhi 110007, India}
\newcommand{\NJU}{{Nanjing~University, Nanjing}}
\newcommand{\Cincinnati}{Department of Physics, University of Cincinnati, Cincinnati, Ohio 45221, USA}
\newcommand{\Rochester}{Department of Physics and Astronomy, University of Rochester, New York 14627 USA}
\newcommand{\Iowa}{Department of Physics and Astronomy, Iowa State University, Ames, Iowa 50011 USA}
\newcommand{\Athens}{Department of Physics, University of Athens, GR-15771 Athens, Greece}
\newcommand{\XJTU}{{Department of Nuclear Science and Technology, School of Energy and Power Engineering, Xi'an Jiaotong University, Xi'an}}
\newcommand{\Charles}{{Charles~University, Faculty~of~Mathematics~and~Physics, Prague, Czech Republic}} 
\newcommand{\Yale}{{Wright~Laboratory and Department~of~Physics, Yale~University, New~Haven, Connecticut 06520, USA}} 
\newcommand{\NTU}{{Department of Physics, National~Taiwan~University, Taipei}}
\newcommand{\PennU}{Department of Physics and Astronomy, University of Pennsylvania, Philadelphia, Pennsylvania 19104, USA}
\newcommand{\CdF}{APC -- Universit\'{e} Paris 7 Denis Diderot, 10, rue Alice Domon et L\'{e}onie Duquet, F-75205 Paris Cedex 13, France}
\newcommand{\Oxford}{Subdepartment of Particle Physics, University of Oxford, Oxford OX1 3RH, United Kingdom}
\newcommand{\LANL}{Los Alamos National Laboratory, Los Alamos, New Mexico 87545, USA}
\newcommand{\Caltech}{Lauritsen Laboratory, California Institute of Technology, Pasadena, California 91125, USA}
\newcommand{\Alabama}{Department of Physics and Astronomy, University of Alabama, Tuscaloosa, Alabama 35487, USA}
\newcommand{\Houston}{Department of Physics, University of Houston, Houston, Texas 77204, USA}
\newcommand{\SZU}{{Shenzhen~University, Shenzhen}}
\newcommand{\LLL}{Lawrence Livermore National Laboratory, Livermore, California 94550, USA}
\newcommand{\PennState}{Department of Physics, Pennsylvania State University, State College, Pennsylvania 16802, USA}
\newcommand{\Washington}{Physics Department, Western Washington University, Bellingham, Washington 98225, USA}
\newcommand{\HKU}{{Department of Physics, The~University~of~Hong~Kong, Pokfulam, Hong~Kong}}
\newcommand{\TsingHua}{{Department~of~Engineering~Physics, Tsinghua~University, Beijing}}
\newcommand{\UNICAMP}{Universidade Estadual de Campinas, IFGW, CP 6165, 13083-970, Campinas, SP, Brazil}
\newcommand{\Indiana}{Indiana University, Bloomington, Indiana 47405, USA}
\newcommand{\deceased}{Deceased.}
\newcommand{\Otterbein}{Otterbein University, Westerville, Ohio 43081, USA}
\newcommand{\Dallas}{Department of Physics, University of Dallas, Irving, Texas 75062, USA}
\newcommand{\UIUC}{{Department of Physics, University~of~Illinois~at~Urbana-Champaign, Urbana, Illinois 61801, USA}}
\newcommand{\RAL}{Rutherford Appleton Laboratory, Science and Technology Facilities Council, Didcot, OX11 0QX, United Kingdom}
\newcommand{\ZSU}{{Sun Yat-Sen (Zhongshan) University, Guangzhou}}
\newcommand{\BCC}{{Now at Department of Chemistry and Chemical Technology, Bronx Community College, Bronx, New York  10453, USA}} 
\newcommand{\Manchester}{Department of Physics and Astronomy, University of Manchester, Manchester M13 9PL, United Kingdom}
\newcommand{\GEHealth}{GE Healthcare, Florence South Carolina 29501, USA}
\newcommand{\SDU}{{Shandong~University, Jinan}}
\newcommand{\Crookston}{Math, Science and Technology Department, University of Minnesota -- Crookston, Crookston, Minnesota 56716, USA}
\newcommand{\RoyalH}{Physics Department, Royal Holloway, University of London, Egham, Surrey, TW20 0EX, United Kingdom}
\newcommand{\LosAlamos}{Los Alamos National Laboratory, Los Alamos, New Mexico 87545, USA}
\newcommand{\CQU}{{Chongqing University, Chongqing}} 
\newcommand{\IIT}{Department of Physics, Illinois Institute of Technology, Chicago, Illinois 60616, USA}
\newcommand{\Cleveland}{Cleveland Clinic, Cleveland, Ohio 44195, USA}
\newcommand{\ITEP}{High Energy Experimental Physics Department, ITEP, B. Cheremushkinskaya, 25, 117218 Moscow, Russia}
\newcommand{\TechX}{Tech-X Corporation, Boulder, Colorado 80303, USA}
\newcommand{\Harvard}{Department of Physics, Harvard University, Cambridge, Massachusetts 02138, USA}
\newcommand{\Sussex}{Department of Physics and Astronomy, University of Sussex, Falmer, Brighton BN1 9QH, United Kingdom}
\newcommand{\Benedictine}{Physics Department, Benedictine University, Lisle, Illinois 60532, USA}
\newcommand{\Lebedev}{Nuclear Physics Department, Lebedev Physical Institute, Leninsky Prospect 53, 119991 Moscow, Russia}
\newcommand{\Wisconsin}{Physics Department, University of Wisconsin, Madison, Wisconsin 53706, USA}
\newcommand{\Cambridge}{Cavendish Laboratory, University of Cambridge, Cambridge CB3 0HE, United Kingdom}
\newcommand{\BNL}{Brookhaven National Laboratory, Upton, New York 11973, USA}
\newcommand{\NUDT}{{College of Electronic Science and Engineering, National University of Defense Technology, Changsha}} 
\newcommand{\BNU}{{Beijing~Normal~University, Beijing}}
\newcommand{\NanKai}{{School of Physics, Nankai~University, Tianjin}}
\newcommand{\SDakota}{South Dakota School of Mines and Technology, Rapid City, South Dakota 57701, USA}
\newcommand{\DGUT}{{Dongguan~University~of~Technology, Dongguan}}
\newcommand{\UC}{{Department of Physics, University~of~Cincinnati, Cincinnati, Ohio 45221, USA}}
\newcommand{\UFG}{Instituto de F\'{i}sica, Universidade Federal de Goi\'{a}s, 74690-900, Goi\^{a}nia, GO, Brazil}
\newcommand{\IHEP}{Institute for High Energy Physics, Protvino, Moscow Region RU-140284, Russia}
\newcommand{\Princeton}{{Joseph Henry Laboratories, Princeton~University, Princeton, New~Jersey 08544, USA}}
\newcommand{\Carolina}{Department of Physics and Astronomy, University of South Carolina, Columbia, South Carolina 29208, USA}
\newcommand{\Siena}{{Siena~College, Loudonville, New York  12211, USA}}
\newcommand{\Tufts}{Physics Department, Tufts University, Medford, Massachusetts 02155, USA}
\newcommand{\Warsaw}{Department of Physics, University of Warsaw, PL-02-093 Warsaw, Poland}
\newcommand{\UCB}{{Department of Physics, University~of~California, Berkeley, California  94720, USA}}
\newcommand{\UCL}{Department of Physics and Astronomy, University College London, London WC1E 6BT, United Kingdom}
\newcommand{\TempleUniversity}{{Department~of~Physics, College~of~Science~and~Technology, Temple~University, Philadelphia, Pennsylvania  19122, USA}}
\newcommand{\UCI}{{Department of Physics and Astronomy, University of California, Irvine, California 92697, USA}} 
\newcommand{\VirginiaTech}{{Center for Neutrino Physics, Virginia~Tech, Blacksburg, Virginia  24061, USA}}
\newcommand{\Texas}{Department of Physics, University of Texas at Austin, Austin, Texas 78712, USA}
\newcommand{\USP}{Instituto de F\'{i}sica, Universidade de S\~{a}o Paulo,  CP 66318, 05315-970, S\~{a}o Paulo, SP, Brazil}
\newcommand{\Pittsburgh}{Department of Physics and Astronomy, University of Pittsburgh, Pittsburgh, Pennsylvania 15260, USA}
\newcommand{\NTUAthens}{Department of Physics, National Tech. University of Athens, GR-15780 Athens, Greece}
\newcommand{\USTC}{{University~of~Science~and~Technology~of~China, Hefei}}
\newcommand{\CIAE}{{China~Institute~of~Atomic~Energy, Beijing}}
\newcommand{\NCEPU}{{North~China~Electric~Power~University, Beijing}}
\newcommand{\FNAL}{Fermi National Accelerator Laboratory, Batavia, Illinois 60510, USA}
\newcommand{\NCTU}{{Institute~of~Physics, National~Chiao-Tung~University, Hsinchu}}
\newcommand{\Duluth}{Department of Physics, University of Minnesota Duluth, Duluth, Minnesota 55812, USA}
\newcommand{\TexasAM}{Physics Department, Texas A\&M University, College Station, Texas 77843, USA}
\newcommand{\Berkeley}{Lawrence Berkeley National Laboratory, Berkeley, California, 94720 USA}
\newcommand{\HolyCross}{Holy Cross College, Notre Dame, Indiana 46556, USA}
\newcommand{\MIT}{Lincoln Laboratory, Massachusetts Institute of Technology, Lexington, Massachusetts 02420, USA}
\newcommand{\CGNPG}{{China General Nuclear Power Group, Shenzhen}}
\newcommand{\ANL}{Argonne National Laboratory, Argonne, Illinois 60439, USA}
\newcommand{\Dubna}{{Joint~Institute~for~Nuclear~Research, Dubna, Moscow~Region, Russia}}
\newcommand{\CUHK}{{Chinese~University~of~Hong~Kong, Hong~Kong}}
\newcommand{\CNIHEP}{{Institute~of~High~Energy~Physics, Beijing}}
\newcommand{\Stanford}{Department of Physics, Stanford University, Stanford, California 94305, USA}
\newcommand{\Lancaster}{Lancaster University, Lancaster, LA1 4YB, United Kingdom}
\newcommand{\Ohio}{Center for Cosmology and Astro Particle Physics, Ohio State University, Columbus, Ohio 43210 USA}
\newcommand{\StJohnFisher}{Physics Department, St. John Fisher College, Rochester, New York 14618 USA}
\newcommand{\CUC}{{Instituto de F\'isica, Pontificia Universidad Cat\'olica de Chile, Santiago, Chile}} 
\newcommand{\WandM}{Department of Physics, College of William \& Mary, Williamsburg, Virginia 23187, USA}
\author{P.~Adamson\ensuremath{^{\mu}}}
\affiliation{\FNAL} 
\author{F.~P.~An\ensuremath{^{\delta}}}
\affiliation{\ECUST} 
\author{I.~Anghel\ensuremath{^{\mu}}}
\affiliation{\Iowa} 
\author{A.~Aurisano\ensuremath{^{\mu}}}
\affiliation{\Cincinnati} 
\author{A.~B.~Balantekin\ensuremath{^{\delta}}}
\affiliation{\Wisconsin} 
\author{H.~R.~Band\ensuremath{^{\delta}}}
\affiliation{\Yale} 
\author{G.~Barr\ensuremath{^{\mu}}}
\affiliation{\Oxford} 
\author{M.~Bishai\ensuremath{^{\delta}}}
\affiliation{\BNL} 
\author{A.~Blake\ensuremath{^{\mu}}}
\affiliation{\Cambridge}\affiliation{\Lancaster} 
\author{S.~Blyth\ensuremath{^{\delta}}}
\affiliation{\NTU} 
% ...
\collaboration{\ensuremath{^{\delta}}Daya Bay Collaboration}\noaffiliation
\collaboration{\ensuremath{^{\mu}}MINOS+ Collaboration}\noaffiliation

\begin{abstract}
Searches for electron antineutrino, muon neutrino, and muon antineutrino disappearance driven by sterile neutrino mixing have been carried out by the Daya Bay and MINOS+ collaborations. This Letter presents the combined results of these searches, along with exclusion results from the Bugey-3 reactor experiment, framed in a minimally extended four-neutrino scenario. Significantly improved constraints on the $\theta_{\mu e}$ mixing angle are derived that constitute the most stringent limits to date over five orders of magnitude in the sterile mass-squared splitting $\Delta m^2_{41}$, excluding the 90\% C.L. sterile-neutrino parameter space allowed by the LSND and MiniBooNE observations at 90\%~CL$_s$ for $\Delta m^2_{41}<5\,$eV$^2$.  
Furthermore, the LSND and MiniBooNE 99\%~C.L. allowed regions are excluded at 99\%~CL$_s$ for $\Delta m^2_{41}$~$<$~1.2~eV$^2$.

\end{abstract}

\maketitle
%
Since the discovery of neutrino oscillations two decades ago~\cite{Fukuda:1998mi,Ahmad:2001an}, the progress achieved allows more rigorous tests than ever to be conducted. The wealth of experimental data to date overwhelmingly demonstrates that the weak and mass eigenstates of the neutrino mix. 

Most of the measurements made so far with solar, atmospheric, reactor and accelerator neutrinos~\cite{snolatest,SKsolarlatest,SKlatest,icecubelatest,kamlandlatest,dayabaylatest,renolatest,DClatest,t2klatest,novalatest,minos2013separated,minoslatest,operalatest} can be fully explained with three neutrino states that mix as described by the PMNS formalism~\cite{Pontecorvo:1957cp,Pontecorvo:1967fh,Maki:1962mu}. There are, however, some experimental observations that cannot be accommodated in the three-neutrino mixing model, such as the excess of electron-like events in a muon (anti)neutrino beam 
observed over short baselines by the Liquid Scintillator Neutrino Detector (LSND)~\cite{Aguilar:2001ty} and MiniBooNE~\cite{Aguilar-Arevalo:2013pmq,Aguilar-Arevalo:2018gpe} experiments. 
These observations may be explained by mixing with at least one additional fourth neutrino state with mass-squared splitting $\Delta m^2_{41} \gg |\Delta m^2_{32}|$, where the $\Delta m^2_{ji}=m^2_j - m^2_i$ represent neutrino mass-squared differences and $m_i$ is the mass of the $i$-th mass eigenstate. 
The addition of such states, a natural occurrence in many extensions of the Standard Model that incorporate neutrino masses, results in new neutrino states that are commonly deemed to be sterile in accordance with the tight constraints from precision electroweak measurements~\cite{PDG2014,ALEPH:2005ab} on the number of neutrinos that couple to the $Z$ boson.
The far-reaching implications of sterile neutrinos in particle physics and cosmology makes their possible existence one of the key questions in physics. 

Sterile neutrinos could be detected in oscillation experiments as a deviation from the standard three-neutrino oscillation behavior if they mix with the three active neutrinos. In 2016, the Daya Bay and MINOS experiments reported limits on active-to-sterile oscillations obtained by combining the results of their electron antineutrino and muon (anti)neutrino disappearance measurements, respectively~\cite{MINOSDayaBay2016}, with those from the Bugey-3 experiment~\cite{bugey-3:paper}. This Letter presents significantly improved limits obtained by utilizing a data set with roughly twice the exposure in the case of Daya Bay~\cite{dyb_1230days}, and by adding $5.80\times10^{20}$\ protons-on-target (POT) of MINOS+ data, recorded with the medium-energy configuration of the NuMI beam~\cite{Adamson:2015dkw}, to the full MINOS data sample~\cite{2018minosplussterile}. Some key systematic uncertainties are reduced in the case of Daya Bay, and a new two-detector fit technique is employed for MINOS and MINOS+. The resulting limits provide leading constraints on possible mixing between active and sterile neutrinos, and can be used to examine the sterile neutrino interpretation of the appearance claims made by the LSND and MiniBooNE experiments in a way that is independent of CP violation and mass-ordering effects. 

The results of the combined analysis presented in this Letter are interpreted within the framework of a 3+1 model, which includes one new mass eigenstate and one sterile weak eigenstate in addition to the three known mass eigenstates and active neutrino flavors. We parameterize the extended 4$\times$4 unitary matrix $U$ describing mixing between weak and mass eigenstates following Ref.~\cite{Harari:1986xf}, 
and the expressions for the elements of $U$ that are relevant to this work become
\begin{eqnarray}
\begin{split}
|U_{e3}|^2 =& \,\,\cos^2 \theta_{14} \sin^2\theta_{13}, \\
|U_{e4}|^2 =& \,\,\sin^2\theta_{14}, \\
|U_{\mu4}|^2 =& \,\,\sin^2\theta_{24}\cos^2\theta_{14}. 
\end{split}
\end{eqnarray}
Under the assumption of neutrino-antineutrino invariance, in the $\Delta m^2_{41}$ $\gg$ $|\Delta m^2_{31}|$ approximation for Daya Bay and Bugey-3 baselines, the survival probability of electron antineutrinos with energy $E$ after traveling a distance $L$ approximates to 
\begin{eqnarray}
P_{\overline\nu_e\rightarrow\overline\nu_e} &\approx& 
1-4|U_{e4}|^2\left(1-|U_{e4}|\right)^2\sin^2\left(\frac{\Delta m^2_{41}L}{4E}\right) \nonumber \\
&-& 4|U_{e3}|^2\left(1-|U_{e3}|^2\right)\sin^2\left(\frac{\Delta m^2_{31}L}{4E}\right), 
\label{eq:PeeFullU}
\end{eqnarray}
which yields the following $\sin^2 2\theta_{14}$-dependent expression: 
\begin{eqnarray}
P_{\overline\nu_e\rightarrow\overline\nu_e} &\approx& 
1-\sin^2 2\theta_{14} \sin^2\left(\frac{\Delta m^2_{41}L}{4E}\right) \nonumber \\
&-& \sin^2 2\theta_{13} \sin^2\left(\frac{\Delta m^2_{31}L}{4E}\right). 
\label{eq:PeeFullU_Pre}
\end{eqnarray}
Long-baseline experiments like MINOS and MINOS+ constrain $\sin^2\theta_{24}$ by looking for muon neutrino and antineutrino disappearance, for which we can approximate the survival probability as
\begin{eqnarray}
P_{\overset{(-)}\nu\!\!_\mu\rightarrow\overset{(-)}\nu\!\!_\mu}\approx&1&-\sin^{2}2\theta_{23}\cos2\theta_{24}\sin^{2}\left(\frac{\Delta m^{2}_{31} L }{4E}\right) \nonumber \\
&-&\sin^{2}2\theta_{24}\sin^{2}\left(\frac{\Delta m^{2}_{41}L}{4E}\right). \label{eq:probdisapnumu}
\end{eqnarray}
In addition, long-baseline experiments can also look for deficits of neutral-current (NC) neutrino interactions between the Near and Far detectors, approximately described by
\begin{eqnarray}
 P_{\textrm{NC}} &=& 1 - P\left(\nu_\mu\rightarrow\nu_s\right) \nonumber \\
&\approx& 1 -\cos^{4}\theta_{14}\cos^{2}\theta_{34}\sin^{2}2\theta_{24}\sin^{2}\left(\frac{\Delta m^{2}_{41}L}{4E}\right) \nonumber \\
& &- \sin^{2}\theta_{34}\sin^{2}2\theta_{23}\sin^{2}\left(\frac{\Delta m^{2}_{31} L }{4E}\right) \\
& &+\frac{1}{2}\sin\delta_{24}\sin\theta_{24}\sin2\theta_{34}\sin2\theta_{23}\sin\left(\frac{\Delta m^{2}_{31} L }{2E}\right). \nonumber
\end{eqnarray}
Besides sensitivity to both $\theta_{24}$ and $\Delta m^{2}_{41}$, the NC channel provides sensitivity to $\theta_{14}$, $\theta_{34}$ and $\delta_{24}$.
Sterile (anti)neutrino-driven muon to electron (anti)neutrino appearance at short baselines has been advanced as a possible explanation of the LSND and MiniBooNE results. This appearance probability is described by
\begin{align}
P_{\overset{(-)}\nu\!\!_\mu\rightarrow\overset{(-)}\nu\!\!_e}^{SBL} = & \,4|U_{e4}|^2|U_{\mu 4}|^2\sin^2\left(\frac{\Delta m^2_{41}L}{4E}\right), \label{eq:PmueU}
\end{align}
where
\begin{align}
4|U_{e4}|^2|U_{\mu4}|^2 = \sin^22\theta_{14}\sin^2\theta_{24} \equiv \sin^22\theta_{\mu e}.
\label{eq:sin22tmue}
\end{align}
Therefore, electron antineutrino disappearance constraints from reactors on $\sin^22\theta_{14}$, combined with muon neutrino and antineutrino disappearance constraints from long-baseline experiments on $\sin^2\theta_{24}$, can place stringent constraints on the quadratically-suppressed electron neutrino or antineutrino appearance described by $\sin^22\theta_{\mu e}$ within the framework of the 3+1 model~\cite{Gariazzo:2015rra}. While Eqs.~\ref{eq:PeeFullU_Pre} and \ref{eq:probdisapnumu} show  leading terms to illustrate the general behavior of the oscillation probabilities, exact formulae of the full survival probabilities are used in the analyses reported in this Letter.

The Daya Bay reactor antineutrino experiment consists of eight identically-designed antineutrino detectors (ADs) placed in three underground experimental halls (EHs) at different distances from three pairs of 2.9~GW$_{\mathrm{th}}$ nuclear reactors in the southeast of China. The two near halls, EH1 and EH2, house two ADs each and have flux-averaged baselines on the order of 550~m. The far hall, EH3, houses four ADs and has a flux-averaged baseline around 1600~m. 
The overburdens of EH1, EH2 and EH3 are 250, 265, and 860 meters-water-equivalent, respectively. Electron antineutrinos are detected via the inverse beta decay (IBD) reaction, $\bar{\nu}_e + p \rightarrow e^{+} + n$, whose two products are visible in the ADs. Further details about the Daya Bay experiment can be found in Ref.~\cite{DayaBay:detectors}.

Daya Bay's unique configuration with multiple baselines makes it well suited to search for sterile neutrino mixing. A relative comparison of the flux and spectral shape of reactor antineutrinos observed in the EHs at different baselines provides most of the sensitivity to sterile neutrino oscillations in the $10^{-3}$~eV$^2$ $\lesssim |\Delta m^2_{41}| \lesssim 0.3$~eV$^2$ region. For $|\Delta m^2_{41}| \gtrsim 0.3$~eV$^2$, the oscillations are too fast to be resolved by the detectors, and the sensitivity arises primarily from comparing the measured flux with the expectation. The uncertainty in the expected reactor antineutrino flux is conservatively set to 5\% as motivated by recent re-evaluations in light of the so-called reactor antineutrino anomaly~\cite{Hayes:2013wra,Vogel:2016ted}.

A new search for light sterile neutrino mixing was performed at Daya Bay with a data set acquired over 1230 days. This represents a factor of $\sim$2 increase in exposure over the previous result~\cite{dyb_6+8_sterile}. The analysis of this data set incorporates other improvements, such as a more precise background assessment, the inclusion of a time-dependent correction for spatial nonuniformity within each AD, and a reduction in the relative detection efficiency uncertainty to 0.2\%, which is the dominant source of systematic error. The IBD selection, background rejection, and assessment of systematic uncertainties for this data set are described in detail in Ref.~\cite{dyb_1230days}. The normal mass ordering is assumed for $\Delta m^2_{31}$ and $\Delta m^2_{41}$. The results reported here are largely insensitive to this choice.

The same two complementary methods applied in previous sterile neutrino searches at Daya Bay~\cite{DYB:sterile6AD,dyb_6+8_sterile} are used to set the exclusion limits in the $(\Delta m^2_{41}, \sin^2 2\theta_{14})$ parameter space. The first one is based on a purely relative comparison between the near and the far data, and relies on the frequentist approach proposed by Feldman and Cousins to determine the exclusion limits~\cite{Feldman:1997qc}. The second one uses the predicted antineutrino spectra to simultaneously fit the observations in the three halls, and uses the CL$_s$ statistical method~\cite{Read:2002hq,Junk1999435} to set the limits. 

The CL$_s$ method is a two-hypothesis test, here used to discriminate between the three-neutrino ($3\nu$) and four-neutrino ($4\nu$) scenarios where each combination of ($\Delta m^2_{41}$, $\sin^2 2\theta_{14}$) is treated as a separate $4\nu$ scenario. 
We define the test statistic $\Delta \chi^2 = \chi^2_{4\nu} - \chi^2_{3\nu}$, where $\chi^2_{3\nu}$ is the $\chi^2$ resulting from a fit to the $3\nu$ hypothesis (with free $\theta_{13}$) and $\chi^2_{4\nu}$ is the $\chi^2$ from a fit to the $4\nu$ hypothesis (with free $\theta_{13}$, and $\Delta m^2_{41}$ and $\theta_{14}$ set to the corresponding $4\nu$ scenario under consideration).
Other parameters, namely $\sin^2 2\theta_{12}$, $\Delta m^2_{21}$ and $|\Delta m^2_{32}|$, are constrained using external data~\cite{PDG2014}. 
We produce a $\Delta \chi^2_{3\nu}$ distribution by fitting simulated pseudo-experiments with $\Delta m^2_{41} = \sin^2 2\theta_{14} =0$ and $\theta_{13}$ fixed to the best-fit value in the data. The same is done to construct a $\Delta \chi^2_{4\nu}$ distribution for every point in the ($\Delta m^2_{41}$, $\sin^2 2\theta_{14}$) parameter space.
Since these distributions are normally distributed, we estimate their mean and variance from Asimov data sets~\cite{CLsMethod}, greatly reducing the amount of computation needed. For each point in ($\Delta m^2_{41}$, $\sin^2 2\theta_{14}$) the observed $\Delta \chi^2_\mathrm{obs}$ is compared to the $\Delta \chi^2_{3\nu}$ and $\Delta \chi^2_{4\nu}$ distributions in order to obtain the corresponding p-values.
The CL$_s$ statistic is defined by
\begin{equation}
\mathrm{CL_{\it s}} = \frac{1-p_{4\nu}}{1-p_{3\nu}},
\end{equation}
where $p_{\mathrm{H}}$ is the p-value for hypothesis H. The 90\% exclusion contour is obtained by requiring CL$_s \leq 0.1$. 

\begin{figure}[!htbp]
\includegraphics[width=\columnwidth]{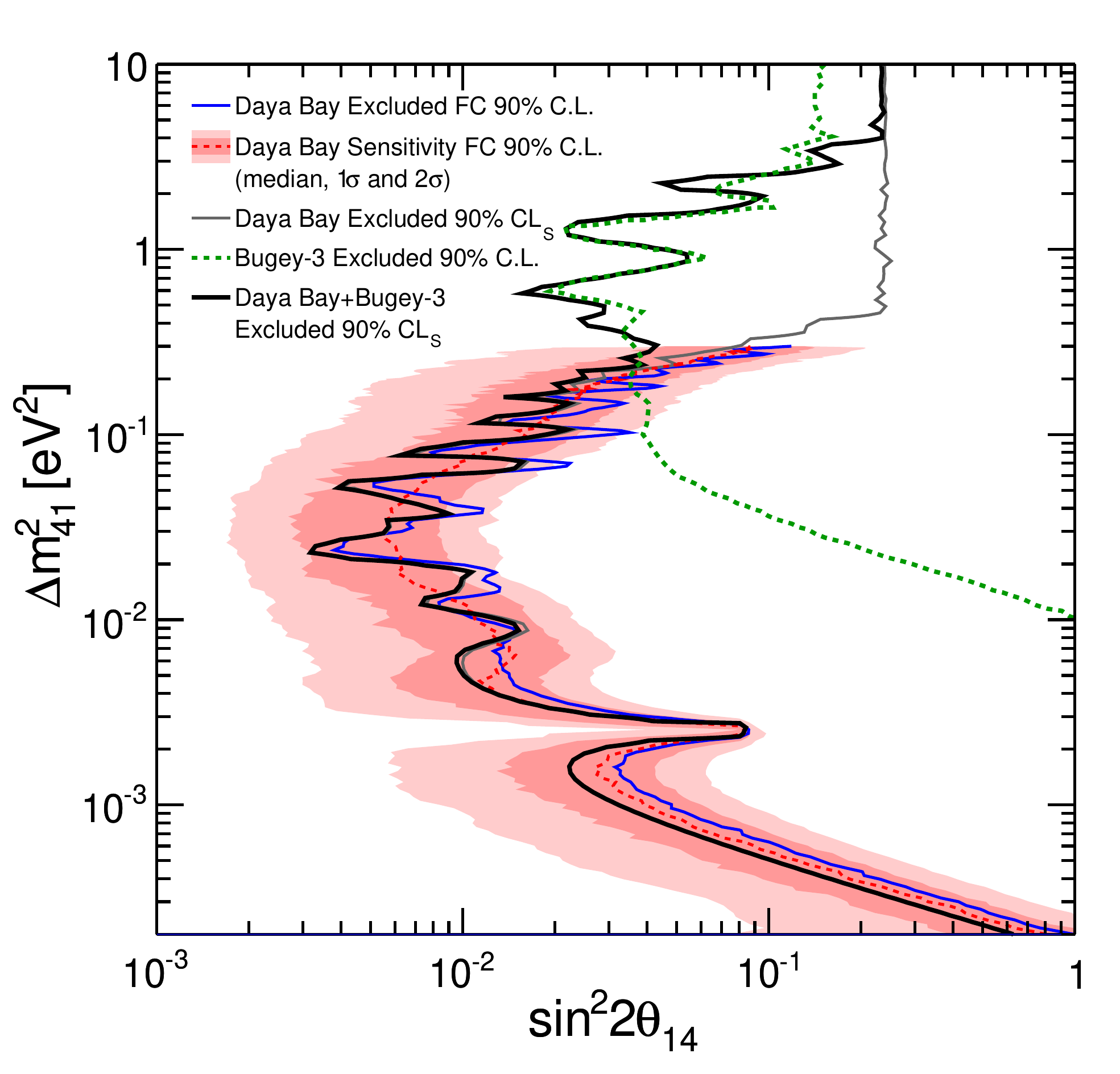} 
\caption{\label{fig:DYBBGY} The Feldman-Cousins (FC) exclusion region at 90\%~C.L. from the analysis of 1230 days of Daya Bay data is shown as the solid blue line. The 90\%~C.L. median sensitivity is shown as the dashed red line, along with $1\sigma$ and $2\sigma$ bands. The excluded region for the original Bugey-3 limit with the raster scan technique is shown in green, while the resulting \cls\ contour from Daya Bay and its combination with the reproduced Bugey-3 results with adjusted fluxes are shown in grey and black, respectively. The regions to the right of the curves are excluded at the 90\%~\cls\ or 90\%~C.L. }
\end{figure}

As seen in Fig.~\ref{fig:DYBBGY}, consistent results are obtained by the two methods. It has been shown that the CL$_s$ approach can yield more stringent contours than the Feldman-Cousins approach with null data sets~\cite{CLsMethod}. Moreover, a study using a very large number of simulated experiments found that the purely relative near-far comparison method that is used to produce the Feldman-Cousins contours had slightly lower sensitivity in the $\Delta m^2_{41}\lesssim 2\times10^{-3}~\mathrm{eV}^2$ region than the method where the near and far observations are fit simultaneously. This study also found that the two methods can react slightly differently to statistical fluctuations. The small differences observed in Fig.~\ref{fig:DYBBGY} 
are thus well within expectation.

A CL$_s$-based analysis is also applied to the published data from the Bugey-3 experiment~\cite{bugey-3:paper}. This reactor experiment operated at shorter ($<$100~m) baselines, allowing it to provide valuable constraints on sterile neutrino mixing from electron antineutrino disappearance for higher values of $\Delta m^2_{41}$ compared to Daya Bay. The same methodology detailed in Ref.~\cite{MINOSDayaBay2016} was followed to generate the exclusion contour for Bugey-3. The main adjustments made with respect to the original Bugey-3 analysis were: (i) the use of the Gaussian CL$_s$ method, instead of the raster scan technique; (ii) the use of an updated neutron lifetime in the IBD cross-section calculation; and (iii) the use of the Huber+Mueller~\cite{Huber:2011wv,Mueller:2011nm} model instead of the original ILL+Vogel model~\cite{ILL,vogel} to make the flux prediction at the different baselines. The reproduced contour is very similar to the one published originally by the Bugey-3 collaboration, shown in Fig.~\ref{fig:DYBBGY}. 


The MINOS and MINOS+ experiments used two detectors placed on the NuMI beam axis, the Near Detector (ND), located $1.04\,$km downstream from the production target at Fermilab at a depth of 225 meters-water-equivalent, and the Far Detector (FD), located 734~km further downstream, in the Soudan Underground Laboratory in Minnesota at a depth of 2070 meters-water-equivalent. The detectors were functionally-identical magnetized, tracking, sampling calorimeters composed of steel-scintillator planes read out by multi-anode photomultiplier tubes~\cite{MINOSNIM}. The NuMI neutrino beam is produced by colliding 120\,GeV protons accelerated by the Main Injector complex at Fermilab with a graphite target. The emerging secondary beam of mostly $\pi$ and $K$ mesons is focused by two parabolic electromagnetic horns and allowed to decay in a 675\,m long helium-filled pipe, resulting in a neutrino beam composed predominantly of $\nu_\mu$, with a 1.3\% contamination of $\nu_e$~\cite{ref:numi}. The detectors accumulated a $10.56\times10^{20}$ POT beam exposure during the MINOS neutrino runs, with the observed neutrino energy spectrum peaked at $3\,$GeV. 
In the MINOS+ phase, the detectors sampled a higher-intensity NuMI beam, upgraded as part of the NOvA experiment~\cite{NOvATDR}, with the neutrino energy peaked at 7\,GeV. 
The higher-energy neutrino beam, although less favorable for three-flavor oscillation measurements (for MINOS' baseline and three-neutrino standard oscillations, the muon neutrino disappearance maximum occurs at $E_\nu \approx 1.6$\,GeV), provides greater sensitivity to sterile-induced muon neutrino disappearance by increasing the statistics in regions of $L/E_{\nu}$ where oscillations driven by large mass-squared splittings would occur.
A new search for sterile neutrino mixing using an additional exposure of $5.80\times10^{20}$\,POT of MINOS+ data has been recently published~\cite{2018minosplussterile}. Unlike the previous MINOS analysis that was based on the ratio between the measured neutrino energy spectra in the two detectors (Far-over-Near ratio)~\cite{MINOSSterile2016,Adamson:2011ku, Adamson:2010wi, Adamson:2008jh} and that was limited by the statistical error of the lower-statistics FD sample, the new analysis employs a two-detector fit method, simultaneously fitting the reconstructed neutrino energy spectra in both detectors~\cite{Todd:2018hin}. The new technique exploits the full power of the large ND statistics for $L/E_{\nu}$ regions probed by the ND baseline. 

The analysis employs both the charged-current (CC)~$\nu_\mu$ and the NC data samples from MINOS and MINOS+. The CC $\nu_\mu$ disappearance channel has sensitivity to $\theta_{24}$ and $\Delta m^{2}_{41}$, in addition to the three-flavor oscillation parameters $\Delta m^{2}_{32}$ and $\theta_{23}$. The NC sample adds nontrivial sensitivity to $\theta_{34}$, $\theta_{24}$ and $\Delta m^{2}_{41}$, albeit with a worse energy resolution (due to the missing energy carried by the outgoing final-state neutrino) than in the CC case, as well as lower statistics due to the lower NC interaction cross section. As detailed in Refs.~\cite{2018minosplussterile, Todd:2018hin}, the analysis is approximately independent of the angle $\theta_{14}$ and the phases $\delta_{13}$, $\delta_{14}$, and $\delta_{24}$, so these parameters are all set to zero in the fit. The MINOS and MINOS+ combined search for sterile neutrinos places the most stringent limit to date on the mixing parameter $\sin^2\theta_{24}$ for most values of the sterile neutrino mass-splitting $\Delta m^2_{41} > 10^{-4}\,$eV$^2$. 

\begin{figure}[!htbp]
\includegraphics[width=\columnwidth]{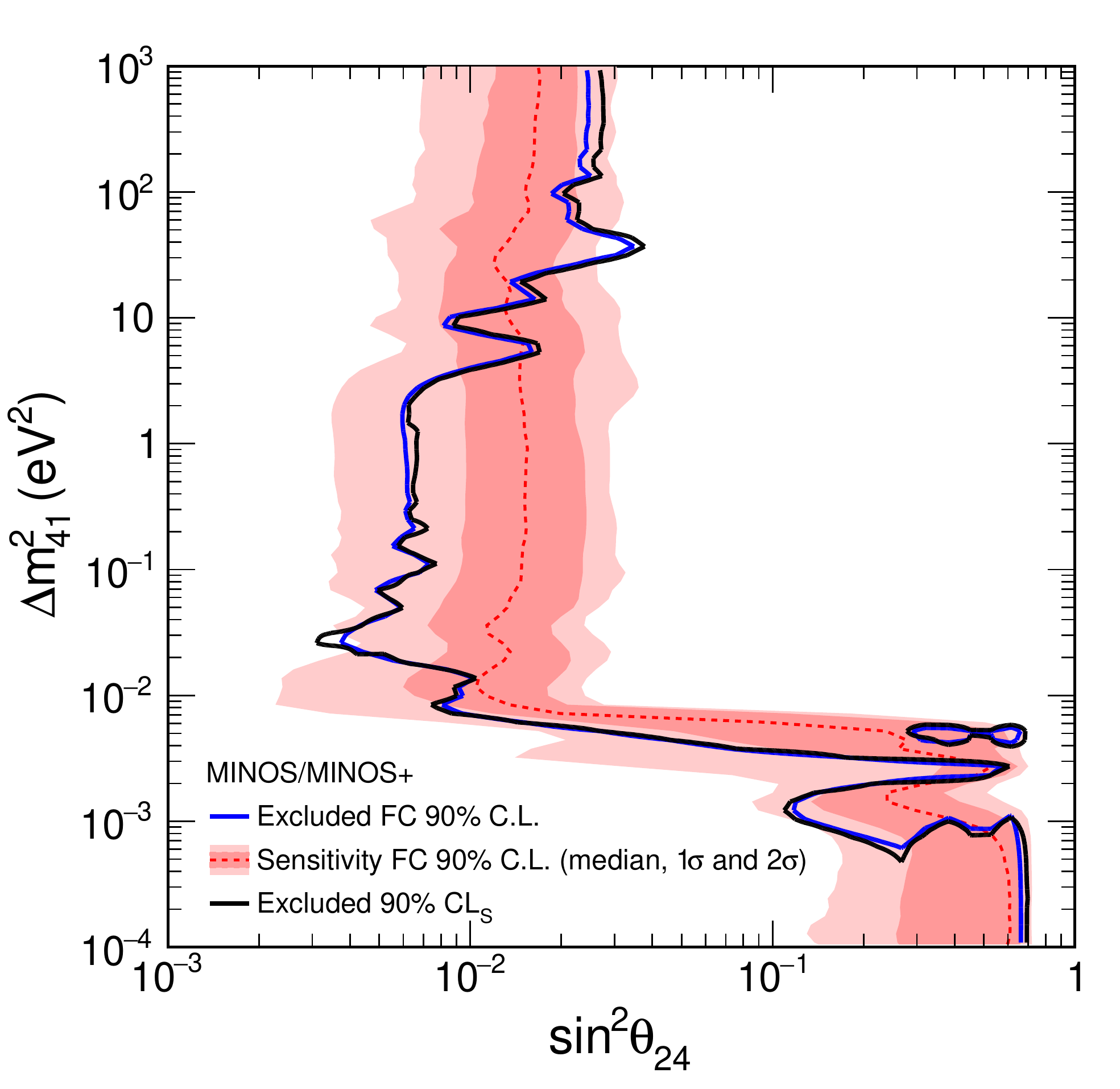}   \caption{\label{fig:minospluscls} Comparison of the MINOS and MINOS+ 90\%~C.L. exclusion contour using the Feldman-Cousins method~\cite{newminossterile} and the \cls\ method. The regions to the right of the curves are excluded at the 90\%~C.L. (\cls). The 90\%~C.L. median sensitivity is shown in red along with the $1\sigma$ and $2\sigma$ bands.}
\end{figure}

Following the same approach used in the first joint analysis by MINOS and Daya Bay~\cite{MINOSDayaBay2016}, the \cls~contours for the new two-detector fit of MINOS and MINOS+ data are obtained using a similar prescription to the one used by Daya Bay, but where the test statistics $\Delta \chi^2_{3\nu}$ and $\Delta \chi^2_{4\nu}$ are approximated by MC simulations of pseudo-experiments without assuming they have Gaussian distributions. The consistency with the published Feldman-Cousins corrected limits is displayed in Fig.~\ref{fig:minospluscls}.
The new MINOS and MINOS+ limits are combined with the Daya Bay and Bugey-3 limits described above to obtain a new improved limit on anomalous $\nu_\mu$ to $\nu_e$ oscillations, as discussed below. 

The disappearance measurements from the three experiments are combined using the same methodology as in Ref.~\cite{MINOSDayaBay2016}. For each fixed value of $\Delta m^2_{41}$, the $\Delta \chi^2_\mathrm{obs}$ value and the $\Delta \chi^2_{3\nu}$ and $\Delta \chi^2_{4\nu}$ distributions for each ($\sin^2 2\theta_{14}$, $\Delta m^2_{41}$) point from the Daya Bay and Bugey-3 combination are paired with those for each ($\sin^2\theta_{24}$, $\Delta m^2_{41}$) point from the MINOS and MINOS+ experiments,  resulting in specific ($\sin^2 2\theta_{\mu e}$, $\Delta m^2_{41}$) combinations according to Eq.~\ref{eq:sin22tmue}. 
Since systematic uncertainties of accelerator and reactor experiments are largely uncorrelated, the combined values of $\Delta \chi^2_\mathrm{obs}$ are obtained by simply summing the corresponding values from the reactor and accelerator experiments. Similarly, the combined $\Delta \chi^2_{3\nu}$ and $\Delta \chi^2_{4\nu}$ distributions are calculated by random sampling the distributions from each experiment and summing. 
Since several different combinations of ($\sin^2 2\theta_{14}$, $\sin^2\theta_{24}$) can yield the same $\sin^2 2\theta_{\mu e}$, the combination with the largest \cls\ value is conservatively selected to be used in the final result. 

\begin{figure}[!htbp]
\includegraphics[width=\columnwidth]{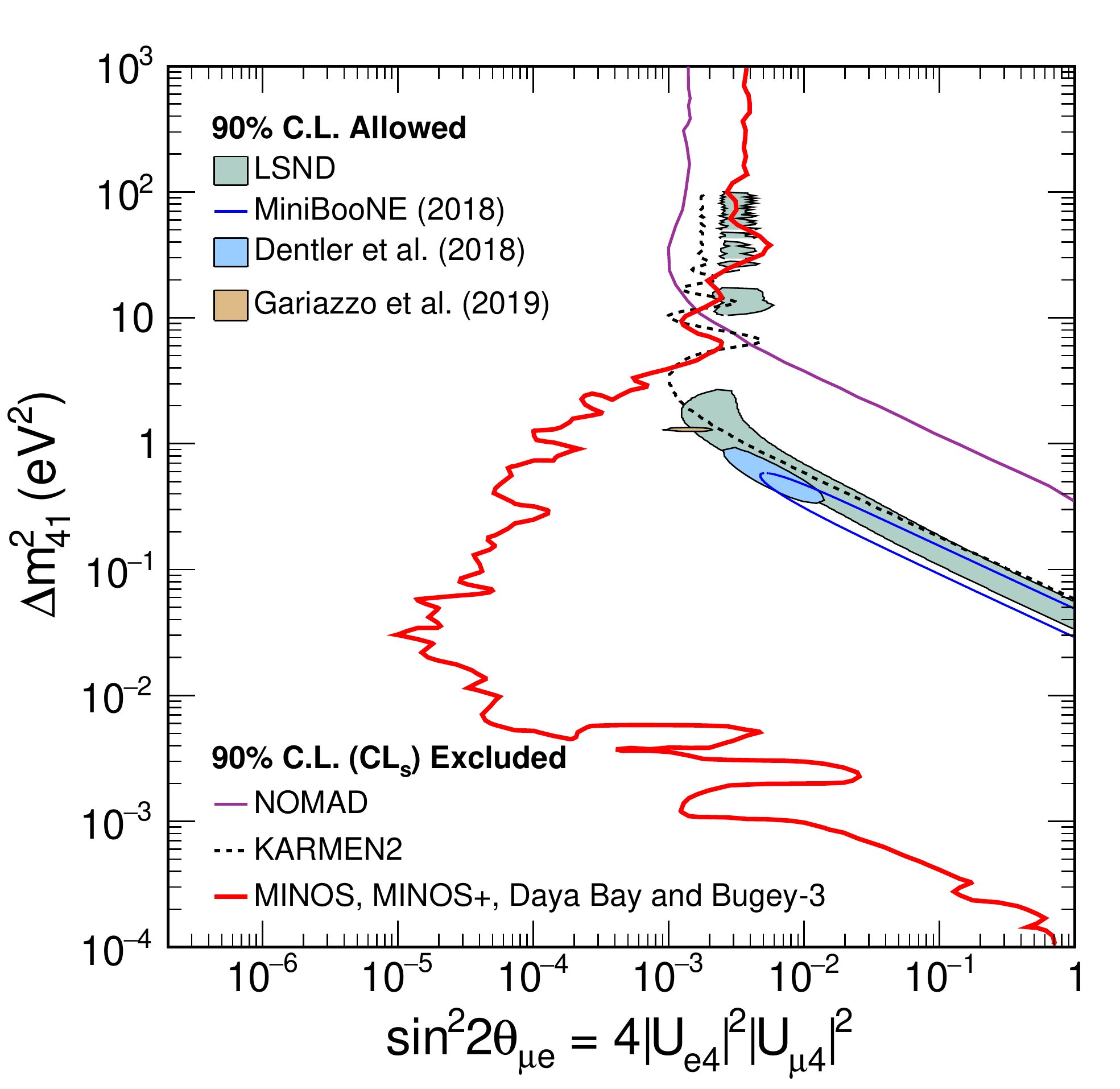}
\caption{\label{fig:combo90cls} Comparison of the MINOS, MINOS+, Daya Bay, and Bugey-3 combined 90\%~\cls\ limit on ${\sin}^22{\theta}_{\mu e}$ to the LSND and MiniBooNE 90\%~C.L. allowed regions. Regions of parameter space to the right of the red contour are excluded. The regions excluded at 90\%~C.L. by the KARMEN2 Collaboration~\cite{Armbruster:2002mp} and
the NOMAD Collaboration~\cite{Astier:2003gs} are also shown. The combined limit also excludes the 90\%~C.L. region allowed by a fit to global data by Gariazzo {\it et al.} where MINOS, MINOS+, Daya Bay, and Bugey-3 are not included~\cite{Gariazzo:2017fdh,Gariazzo:2018mwd}, and the 90\%~C.L. region allowed by a fit to all available appearance data by Dentler {\it et al.}~\cite{Dentler:2018sju} updated with the 2018 MiniBooNE appearance results~\cite{Aguilar-Arevalo:2018gpe}.} 
\end{figure}

\begin{figure}[!htbp]
	\centering
   \includegraphics[width=\columnwidth]{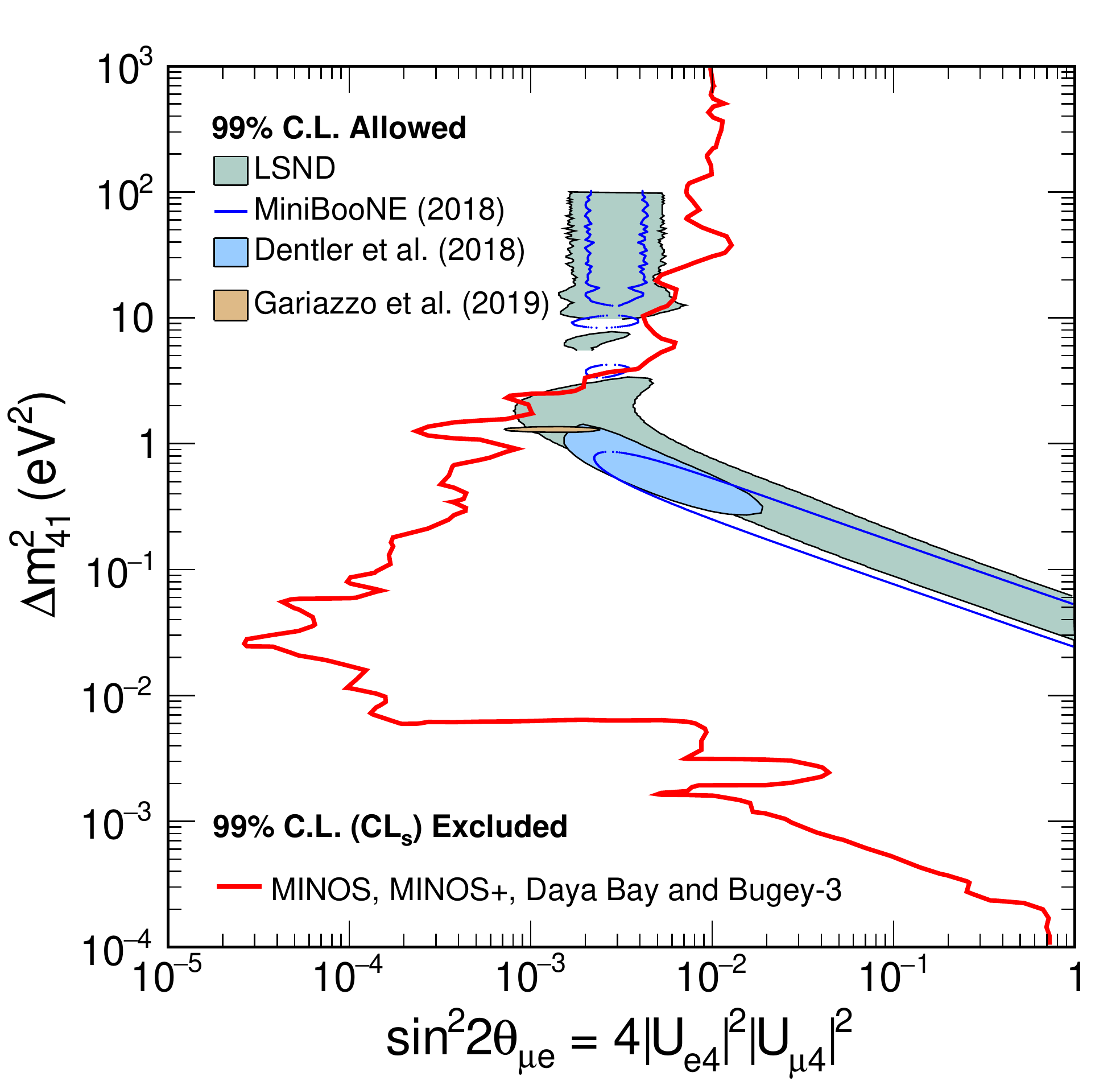}
     \caption{\label{fig:combo99cls} Comparison of the MINOS, MINOS+, Daya Bay, and Bugey-3 combined 99\%~\cls\ limit on ${\sin}^22{\theta}_{\mu e}$ to the LSND and MiniBooNE 99\%~C.L. allowed regions. The limit also excludes the 99\%~C.L. region allowed by a fit to global data by Gariazzo {\it et al.} where MINOS, MINOS+, Daya Bay, and Bugey-3 are not included~\cite{Gariazzo:2017fdh,Gariazzo:2018mwd}, and the 99\%~C.L. region allowed by a fit to all available appearance data by Dentler {\it et al.}~\cite{Dentler:2018sju} updated with the 2018 MiniBooNE appearance results~\cite{Aguilar-Arevalo:2018gpe}.}
\end{figure}

The new combined 90\% and 99\% \cls\ limits from searches for sterile neutrino mixing in MINOS, MINOS+, Daya Bay, and Bugey-3 in the 3+1 neutrino model are shown in Figs.~\ref{fig:combo90cls} and \ref{fig:combo99cls}, respectively. Constraints on the  $\sin^2 2\theta_{\mu e}$ electron (anti)neutrino appearance parameter are provided over 7 orders of magnitude in the sterile mass-squared splitting $\Delta m^2_{41}$. These limits are the world's most stringent over 5 orders of magnitude, for $\Delta m^2_{41} \lesssim 10\,$eV$^2$.

The new constraints exclude the entire 90\%~C.L. allowed regions from LSND and MiniBooNE for $\Delta m^2_{41}<5\,$eV$^2$, with regions at higher values being excluded by NOMAD~\cite{Astier:2003gs}. Further, the 99\%~C.L. allowed regions from LSND and MiniBooNE are excluded for $\Delta m^2_{41}<1.2\,$eV$^2$. The allowed region from a global fit to data from sterile neutrino probes, intentionally excluding MINOS, MINOS+, Daya Bay, and Bugey-3 contributions, computed by the authors of Refs.~\cite{Gariazzo:2017fdh,Gariazzo:2018mwd}, is fully excluded at the 99\%~C.L. The allowed region resulting from a fit to all appearance data, updated by the authors of Ref.~\cite{Dentler:2018sju} to include the MiniBooNE 2018 results~\cite{Aguilar-Arevalo:2018gpe}, is equally strongly excluded. 
The new limits presented here thus significantly increase the tension between pure sterile neutrino mixing explanations of appearance-based indications and the null results from disappearance searches. The sole consideration of additional sterile neutrino states  
cannot resolve this tension, which stems from the non-observation of $\bar{\nu}_e$ and $\pbar{\nu}_{\mu}$ disappearance beyond what is expected from the three-neutrino mixing model.
This inconsistency may be further quantified in additional detector exposures in the process of being analyzed, specifically the last year of MINOS+ data taking, representing an additional sample of similar size to the one used here, as well as over two more years of Daya Bay data. 
\\
\\

We gratefully acknowledge valuable contributions by Carlo Giunti, for supplying a custom fit to global data excluding MINOS, MINOS+, Daya Bay, and Bugey-3 data, and by Mona Dentler and Joachim Kopp, for providing an updated version of a fit to global appearance data including information from the 2018 MiniBooNE appearance results.

The Daya Bay experiment is supported in part by 
the Ministry of Science and Technology of China,
the U.S. Department of Energy,
the Chinese Academy of Sciences,
the CAS Center for Excellence in Particle Physics,
the National Natural Science Foundation of China,
the Guangdong provincial government,
the Shenzhen municipal government,
the China General Nuclear Power Group,
the Research Grants Council of the Hong Kong Special Administrative Region of China,
the Ministry of Education in Taiwan,
the U.S. National Science Foundation,
the Ministry of Education, Youth and Sports of the Czech Republic,
the Joint Institute of Nuclear Research in Dubna, Russia,
the NSFC-RFBR joint research program,
and the National Commission for Scientific and Technological Research of Chile.
We acknowledge Yellow River Engineering Consulting Co., Ltd.\ and China Railway 15th Bureau Group Co., Ltd.\ for building the underground laboratory.
We are grateful for the ongoing cooperation from the China Guangdong Nuclear Power Group and China Light~\&~Power Company.

The MINOS and MINOS+ Collaborations use the resources of the Fermi National Accelerator Laboratory (Fermilab), a U.S. Department of Energy, Office of Science, HEP User Facility. Fermilab is managed by Fermi Research Alliance, LLC (FRA), acting under Contract No. DE-AC02-07CH11359. This work was supported by the U.S. DOE; the United Kingdom STFC, part of UKRI; the U.S. NSF; the State and University of Minnesota; and Brazil's FAPESP, CNPq and CAPES. We thank the personnel of Fermilab's Accelerator and Scientific Computing Divisions and the crew of the Soudan Underground Laboratory for their effort and dedication. We thank the Texas Advanced Computing Center at The University of Texas at Austin for the provision of computing resources. The MINOS and MINOS+ Collaborations acknowledge fruitful cooperation with the Minnesota DNR.

\bibliographystyle{apsrev4-1}
\bibliography{sample}

\end{document}